\documentclass{article}
\usepackage[utf8]{inputenc}
\usepackage{spconf,amsmath,graphicx,mystyle,multirow,booktabs}
\usepackage{subcaption}
\usepackage[hyphens]{url}
\usepackage{hyperref}
\usepackage[hyphenbreaks]{breakurl}

\title{GPU-Accelerated Forward-Backward algorithm with Application to
Lattice-Free MMI}

\name{Lucas Ondel$^{\star}$ \quad Léa-Marie Lam-Yee-Mui$^{\star}$ \quad Martin Kocour$^{\dagger}$
\quad Caio Filippo Corro$^{\star}$ \quad Lukás Burget$^{\dagger}$}

\address{$^{\star}$ LISN, CNRS, Université Paris-Saclay, Orsay, France \\
$^{\dagger}$ Brno University of Technology,  Faculty of Information Technology, Brno, Czech Republic \\
ondel@lisn.fr}

\begin{document}

\ninept
\maketitle

\begin{abstract}
We propose to express the forward-backward algorithm in terms of
operations between sparse matrices in a specific semiring. This new
perspective naturally leads to a GPU-friendly algorithm which is easy
to implement in Julia or any programming languages with native
support of semiring algebra. We use this new implementation to train a
TDNN with the LF-MMI objective function and we compare the training
time of our system with PyChain---a recently introduced C++/CUDA
implementation of the LF-MMI loss. Our implementation is about two
times faster while not having to use any approximation such as the
``leaky-HMM''.

\end{abstract}


\section{Introduction}
\label{sec:intro}
The forward-backward algorithm is a crucial algorithm in speech
recognition. It is used to compute the posterior distribution of state
occupancy in the Expectation-Maximization training approach for
Hidden Markov Model (HMM). Even though deep learning approaches have
superseded the traditional GMM/HMM-based ASR
\cite{morgan1995introduction,xiong2016achieving}, the forward-backward
algorithm is still used to estimate the gradient of two major sequence
discriminative objective functions: Connectionist Temporal Classification
\cite{graves2006connectionist} and Lattice-Free MMMI (LF-MMI)
\cite{povey2016purely,hadian2018end}.

Because state-of-the-art models are trained on GPU, having a fast
and an efficient implementation of these losses (and their gradients)
is essential. The implementation is usually done in C++ and then wrapped
in a Python module to be integrated with popular neural network
libraries \cite{abadi2016tensorflow,paszke2019pytorch}. However, this
practice is far to be satisfactory as the C++ code is usually complex,
difficult to modify and, as we shall see, not necessarily optimal.

In this work, we propose a different approach: we express the
forward-backward algorithms in terms of operations between matrices
living in a particular semiring. This new perspective leads to a
trivial implementation in the Julia language~\cite{Julia-2017} which is
significantly faster than a competitive C++ implementation.  As a
result, our proposed implementation\footnote{
    \url{https://github.com/lucasondel/MarkovModels.jl}, the
forward-backward is implemented in the functions
\texttt{$\alpha$recursion} and \texttt{$\beta$recursion}
} of the forward-backward is just few lines long, easy to read, and
easy to extend by anyone.

This paper is organized in two parts: in Section \ref{sec:algo},
we describe the forward-backward algorithm and its representation
in terms of semiring algebra and in Section \ref{sec:results}, we
conduct our numerical analysis.

Finally, we warmly encourage interested readers to look at the
provided code and the Pluto notebooks\footnote{
   \url{https://github.com/lucasondel/SpeechLab/tree/main/recipes/lfmmi }
}; we have made them with the hope to be accessible by a vast majority.

\section{Algorithm}
\label{sec:algo}
\subsection{Description}

Let $\Matrix{x} = (x_1, x_2, \dots, x_N) $ be a sequence of features and
$\Matrix{z} = (z_1, z_2, \dots, z_n)$ an unobserved sequence sampled
from a discrete-time Markov process defined by the transition
probability $p(z_n | z_{n-1})$. Each $z_n$ takes value in $\{1, \dots, K \}$,
where $1, \dots, K$ are the index of the states of the Markov process.
The forward-backward algorithm \cite{rabiner1989tutorial} calculates
the marginal state posterior distribution:
\begin{align}
    p(z_n | \Matrix{x}) &= \frac{\alpha_{n}(z_n) \beta_{n}(z_n)}{\sum_{z_N}\alpha(z_N)},
\end{align}
where $\alpha$ (the forward pass) and $\beta$ (the backward pass) are
recursive functions defined as:
\begin{align}
    \alpha_n(z_n) &= p(x_n | z_n) \sum_{z_n} p(z_n | z_{n-1}) \alpha_{n-1}(z_{n-1})
        \label{algo:eq:forward}\\
    \beta_n(z_n) &= \sum_{z_{n+1}} p(x_{n+1}|z_{n+1}) p(z_{n+1}|z_n) \beta_{n+1}(z_{n+1})
        \label{algo:eq:backward}.
\end{align}

Whereas the algorithm is simple to state, its implementation can be
quite challenging. First, because it involves multiplication of
probabilities, a naive implementation would quickly underflow. Second,
the state-space is usually very large leading to heavy computations.
However, because it is frequent that $p(z_n|z_{n-1})$ is zero for most
of the pairs $z_n, z_{n-1}$ the amount of ``useful'' operations remains
relatively low.

Thus, an efficient implementation of the forward-backward algorithm should
address these two issues: numerical stability and using the structure
of the transition probabilties to gain speed.

\subsection{Matrix-based implementation}

A convenient way to implement the forward-backward algorithm is to
express \eqref{algo:eq:forward} and \eqref{algo:eq:backward} in terms
of matrix operations.
We introduce the following notation:
\begin{align*}
\mathbf{T} &= \begin{bmatrix}
        p(z_n = 1 | z_{n-1} = 1) & \hdots & p(z_{n-1} = K | z_{n-1} = 1) \\
        \vdots & \ddots & \vdots \\
        p(z_n = 1 | z_{n-1} = K) & \hdots & p(z_n = K | z_{n-1} = K)
    \end{bmatrix} \\
    \mathbf{v}_n &= \begin{bmatrix} p(x_n | z_n = 1) \\ \vdots \\ p(x_n | z_n = K) \end{bmatrix}
        \Matrix{\alpha}_n = \begin{bmatrix}
            \alpha_n(\overbrace{1}^{z_n}) \\ \vdots \\ \alpha_n(K)
            \end{bmatrix}
    \Matrix{\beta}_n =
        \begin{bmatrix} \beta_n(\overbrace{1}^{z_n}) \\ \vdots \\ \beta_n(K) \end{bmatrix},
\end{align*}
and we rewrite the forward-backward algorithm as:
\begin{align}
    \boldsymbol{\alpha}_n &= \mathbf{v}_n \circ ( \mathbf{T}^\top \boldsymbol{\alpha}_{n-1})
        \label{algo:eq:forward_matrix} \\
    \boldsymbol{\beta}_n &= \mathbf{T} (\boldsymbol{\beta}_{n+1} \circ \mathbf{v}_n),
        \label{algo:eq:backward_matrix}
\end{align}
where $\circ$ is the Hadamard (i.e. element-wise) product. Implementing
\eqref{algo:eq:forward_matrix} and \eqref{algo:eq:backward_matrix} is
trivial using any linear algebra library. Another advantage of this
implementation is that it can be easily accelerated with a GPU as
matrix multiplication is a highly optimized operation on such device.
Finally, we can represent the matrix $\Matrix{T}$ as a sparse matrix
and avoid performing unnecessary operations.

However, despite all these benefits, this implementation remains
numerically unstable.

\subsection{Semiring algebra}

In order to keep the advantages of the matrix-based implementation
while solving the numerical stability issue, we propose to express the
algorithm in terms of matrices in the log-semifield\footnote{
A semifield is a semiring for which all the elements but $\bar{0}$
have a multiplicative inverse. Loosely speaking, a semifield is a
semiring with a division operator.
}.

For a matrix $\Matrix{M}$ with non-negative entries we define:
\begin{align}
    \Matrix{M}^{\text{log}} &= \begin{bmatrix}
        \log M_{11} & \log M_{12} & \dots \\
        \log M_{21} & \ddots  & \\
        \vdots & &
    \end{bmatrix}, \\
    M_{ij}^{\text{log}} &\in \mathcal{S}(\mathbb{R}, \oplus, \otimes, \oslash,
        \bar{0}, \bar{1}),
\end{align}
where $\mathcal{S}$ is the log-semifield defined as:
\begin{align}
    a^{\text{log}} \oplus b^{\text{log}} &= \log \big( e^{a^{\text{log}}} + e^{b^{\text{log}}} \big)
        \label{eq:plus} \\
    a^{\text{log}} \otimes b^{\text{log}} &= a^{\text{log}} + b^{\text{log}}
        \label{eq:times} \\
    a^{\text{log}} \oslash b^{\text{log}} &= a^{\text{log}} - b^{\text{log}}
        \label{eq:div} \\
    \bar{0} &= - \infty \label{eq:zero} \\
    \bar{1} &= 0 \label{eq:one}.
\end{align}
Equipped with these new definitions, we express the forward-backward
algorithm in the logarithmic domain:
\begin{align}
    \boldsymbol{\alpha}_n^{\text{log}} &= \mathbf{v}_n^{\text{log}} \circ (
    \mathbf{T}^{\text{log} \top} \boldsymbol{\alpha}_{n-1}^{\text{log}})
        \label{algo:eq:sforw}\\
    \boldsymbol{\beta}_n^{\text{log}} &= \mathbf{T}^{\text{log}}
    (\boldsymbol{\beta}_{n+1}^{\text{log}} \circ \mathbf{v}_n^{\text{log}})
        \label{algo:eq:sback}\\
    \log p(z_n| \mathbf{x}) &= \frac{\alpha_n^{\text{log}}  (z_n) \beta_n^{\text{log}}  (z_n) }{\sum_{z_N} \alpha_N^{\text{log}}  (z_N)} \label{algo:eq:sfb}.
\end{align}
Altogether, \eqref{algo:eq:sforw}, \eqref{algo:eq:sback} and
\eqref{algo:eq:sfb} leads to an implementation of the forward-backward
algorithm which is (i) trivial to implement (if provided with a semiring
algebra API) as it consists of a few matrix multiplications, (ii)
numerically stable as all the computations are in the logarithmic
domain, and (iii) efficient as the matrix $\Matrix{T}$ can be
represented as a sparse matrix storing only the elements different
than $\bar{0}$.

\subsection{Dealing with batches}

Thus far, we have described our new implementation for one sequence
only. However, when training a neural network it is common practice to
use a batch of input sequences to benefit from the GPU parallelization.
Our matrix representation of the algorithm is easily extended to
accommodate for multiple sequences.

In the following, we drop the superscript $^{\text{log}}$ for the
sake of clarity. We define $\Matrix{T}_i$ the transition probabilities
of the $i$th sequence of the batch of size $I$.
$\Matrix{\alpha}_{i,n}$, $\Matrix{\beta}_{i,n}$ and $\Matrix{v}_{i,n}$
are defined similarly. Now, we set:
\begin{align*}
    \Matrix{T} &= \begin{bmatrix} \Matrix{T}_1 & & \\
        &  \ddots & \\
        & & \Matrix{T}_I
    \end{bmatrix} & \Matrix{v}_n &= \begin{bmatrix} \Matrix{v}_{1,n} \\
    \vdots \\
    \Matrix{v}_{I,n}
    \end{bmatrix} \\
    \Matrix{\alpha}_n &= \begin{bmatrix} \Matrix{\alpha}_{1,n} \\
    \vdots \\
    \Matrix{\alpha}_{I,n}
    \end{bmatrix} & \Matrix{\beta}_n &= \begin{bmatrix} \Matrix{\beta}_{1,n} \\
    \vdots \\
    \Matrix{\beta}_{I,n}
    \end{bmatrix},
\end{align*}
and using these new variables in \eqref{algo:eq:sforw},
\eqref{algo:eq:sback} and \eqref{algo:eq:sfb}, we obtain a batch version
of the forward-backward algorithm. This algorithm naturally extends
to batch of sequences of different length by adding to the Markov
process a phony self-looping state marking the end of sequence and
padding $\Matrix{v}$ with $\bar{0}$ appropriately.

\subsection{Using the Julia language}

The algorithm we have described so far is straightforward to implement
but for one difficulty: packages dealing with sparse matrices usually
store the elements that are different from $0$. However, for the
log-semifield, because $\bar{0} \ne 0$ and $\bar{1} = 0$, this would
lead to ignore important values and store many of non-relevant ones.

Implementing in C or C++ a sparse matrix API agnostic to
the semiring is not a trifle. It is perhaps easier to do it in a
scripting language such as Python, but that would lead to poor
performances.  Fortunately, the Julia language provides an elegant
solution to this problem.

Julia is a high-level language with performances comparable
to other statically compiled languages. From a user perspective, it
provides a programming experience close to what a Python programmer
is accustomed to while allowing to write critical code without resorting
to C or C++\footnote{
It is also possible to write GPU kernels directly in Julia as was done
in this work.
}. Importantly for our problem, Julia naturally allows to use arbitrary
semirings \cite{shah2013novel} and, consequently, implementing a
sparse matrix in the log-semifield amounts to write a few lines of
code to implement \eqref{eq:plus}-\eqref{eq:one}. Moreover, Julia
offers a rich landscape for GPU \cite{besard2018juliagpu}
and neural network toolkits \cite{yuret2016knet, innes2018flux}
allowing to integrate any new code with state-of-the-art machine
learning techniques.

\section{Application}
\label{sec:results}
%
%
We now demonstrate the practicality of our algorithm by using it
to build an ASR system trained with the LF-MMI objective function.

\subsection{Objective function}

The LF-MMI objective function \cite{chow1990maximum,povey2016purely,hadian2018end}
for one utterance is defined as:
\begin{align}
    \mathcal{L} &= \log \frac{p(\Matrix{X} | \mathbb{G}_{\text{num}})}{
        p(\Matrix{X} | \mathbb{G}_{\text{den}})},
\end{align}
where $\mathbb{G}_{\text{num}}$ is the \emph{numerator graph}, i.e.
an utterance-specific alignment graph, and $\mathbb{G}_{\text{den}}$
is the \emph{denominator graph}, i.e.  a phonotactic language model.
If we denote $\Matrix{\Phi} = (\Matrix{\phi}_1, \Matrix{\phi}_2, \dots)$
the sequence output by a neural network where we interpret
$\phi_{n,i} =  \log p(\Matrix{x}_n | z_n = i)$, the derivatives of the
loss are given by:
\begin{align}
    \frac{\partial \mathcal{L}}{\partial \phi_{n,i}} &=
        p(z_n = i | \Matrix{X}, \mathbb{G}_{\text{num}}) -
        p(z_n = i | \Matrix{X}, \mathbb{G}_{\text{den}})
        \label{eq:lfmmi_grad}.
\end{align}
In practice, these derivatives are estimated by running the
forward-backward algorithm on the numerator and denominator graphs.

\subsection{Python and Julia recipes}
The baseline is the PyChain \cite{shao2020pychain} package which
implements the LF-MMI objective function integrated with the PyTorch
\cite{paszke2019pytorch} neural network toolkit. \cite{shao2020pychain}
is the latest development of the original ``Kaldi chain model'' and, to
the best of our knowledge, it is currently the most competitive
implementation of the LF-MMI training. We have used the PyChain
recipe\footnote{
    \url{https://github.com/YiwenShaoStephen/pychain_example}
} provided by the authors to prepare and to train
the system.

We compare the PyChain recipe against our Julia-based recipe\footnote{
\url{https://github.com/lucasondel/SpeechLab/tree/main/recipes/lfmmi}.
} that is built on top of our implementation of the forward-backward
algorithm and the KNet \cite{yuret2016knet} neural network Julia
package. Despite that there exists other popular Julia neural network
packages, we elected to use KNet as it is technically the most similar
to PyTorch.

Note that the PyChain implementation of the LF-MMI loss is not
exact: the forward-backward on the denominator is done with the
so-called ``leaky-HMM'' approximation \cite{povey2016purely} which
speeds up the computations at the expense of an approximate result.

\subsection{Datasets}

We use two datasets:
\begin{itemize}
    \item MiniLibrispeech, a subset of the Librispeech corpus
        \cite{panayotov2015librispeech} created for software testing
        purposes. It contains 5 hours and 2 hours of training and
        validation data respectively. Because, it is only for
        ``debugging'', it doesn't have a proper test set and
        we report the WER on the validation set.
    \item the Wall Street Journal (WSJ) corpus \cite{paul1992design},
        where we use the standard subsets for training (si284),
        validating (dev93) and testing eval92)
\end{itemize}

\subsection{Model and Graphs preparation}

Prior training the model with the LF-MMI objective, one needs to
prepare the alignments graphs (i.e. the numerator graphs) and the
n-gram phonotactic language model (i.e. the denominator graph). Our
preparation is identical to the PyChain recipe with one exception:
whereas the PyChain recipe uses only one pronunciation for words
having multiple ones, we use all the pronunciations when building the
numerator graphs. In both recipes, we set the n-gram order of the
phonotactic language model to 3.

The model is a Time-Delay Neural Network (TDNN) \cite{peddinti2015time}
with 5 convolutional layers and a final affine transformation. Each
convolutional layer is composed of the following sequence:
\begin{itemize}
    \item 1-dimensional convolution
    \item batch-normalization
    \item REctified Linear Unit activation
    \item dropout with a dropping probability of 0.2
\end{itemize}
For each convolutional layer, the kernel sizes are $(3, 3, 3, 3, 3)$,
the strides are $(1, 1, 1, 1, 3)$ and the dilations are $(1, 1, 3, 3, 3)$.

The input features to the model are standard 40-dimensional MFCCs
extracted at a 10 ms frame rate. These features are then mean and
variance normalized on a per-speaker basis.  The neural network output
dimension is set to 84---we have 42 phones and each phone is
modelled a with 2-state HMM topology \cite{hadian2018end} resulting
$2\times 42$ emission probabilities.

\subsection{Training}
Our recipe follows exactly the one of PyChain: we use the Adam
optimizer with $\beta_1 = 0.9, \beta_2 = 0.999$ and an initial learning
rate of $10^{-3}$. If there is no improvement of the validation loss
after one epoch, the learning rate is halved.  We also apply a so-called
``curriculum'' training for one epoch, i.e.  we sort in ascending order
the utterances by their durations for the first epoch. For the rest of
the training, we form batches of sequences of similar lengths and we
shuffle the order of the batches. Upon completion of the training, we
select the model with the best validation loss to decode the test data.

We have observed that our Julia-based recipe consumes more memory
and cannot accommodate the same batch size as the PyChain recipe.
To leverage this issue, we divide the batch size $B$ by a factor $F$ and
we update the model after accumulating the gradient for $F$ batches.
In this way, the gradient calculated is the same but memory
requirement is divided by $F$.

\subsection{Results}
\begin{table*}[t]
\centering
\begin{tabular}{ccccc}
    \toprule
    & & & \multicolumn{2}{c}{\textbf{Duration (s)}} \\
    \textbf{Implementation} & \textbf{Device} & \textbf{Leaky-HMM} & \textbf{Numerator} & \textbf{Denominator} \\
    \toprule
    PyChain &  CPU & no & 5.696 & 421.447 \\
    PyChain &  CPU & yes & \textbf{1.212} & \textbf{27.601} \\
    proposed & CPU & no & 3.789 & 226.60 \\
    \midrule
    PyChain & GPU & no & 0.093 & 5.862 \\
    PyChain & GPU & yes & 0.248 & 5.449 \\
    proposed & GPU & no & \textbf{0.058} & \textbf{1.04} \\
    \bottomrule
\end{tabular}
    \caption{Comparison between the PyChain and the proposed
    implementations of the forward-backward algorithm. On GPU, our
    implementation is significantly faster than both PyChain
    implementations.}
\label{results:tab:bm}
\end{table*}
\begin{table}[ht]
\centering
\begin{tabular}{ccccc}
    \toprule
    \textbf{System} & \textbf{Dataset} & $\mathbf{B/F}$ & \textbf{Duration} & \textbf{WER (\%)} \\
    \toprule
    PyChain & MiniLS & 128/1 & 0h42 & 27.17 \\
    proposed & MiniLS  & 64/2 & \textbf{0h22} & \textbf{21.21} \\
    \midrule
    PyChain & WSJ  & 128/1 & 6h48 & 4.74 \\
    proposed & WSJ & 64/2 & \textbf{3h20} & \textbf{4.37} \\
    \bottomrule
\end{tabular}
\caption{Comparison between the PyChain recipe and our Julia-based
    recipe of LF-MMI. $B$ is the batch size and $F$ is the gradient
    update frequency. Note that we report the duration of the neural
    network training, not the total duration of the recipes.}
\label{results:tab:results}
\end{table}
The results reported here were run with a NVIDIA GeForce RTX 2080 Ti GPU.

The final Word Error Rate (WER) evaluated on the test set and the
duration of the neural network training for both recipes are shown in
Table \ref{results:tab:results}. We observe that Julia-based training
drastically outperforms the baseline recipe in term of training
time even though it cannot use the same batch size.
Regarding the WER, whereas differences for the WSJ is small, our recipe
achieves a much better WER on the MiniLibrispeech corpus. Currently,
we do not have a definitive explanation for this improvement. A
potential cause that we will explore in future work is that the
approximation used in PyChain may degrade the performance of the system
when trained on small amount of data.

\subsection{Analysis}

In order to gain further insights about the training speed up observed
in the previous experiment, we compare our proposed implementation of
the forward-backward algorithm with the ones found in PyChain.
PyChain ships two versions of the forward-backward both
implemented in C++: the first one calculates the exact computations in
the logarithmic domain whereas the second runs in the probability domain
but uses a scaling factor and the ``leaky-HMM'' approximation
to overcome numerical stability issues.

We run our benchmark on two different graphs reflecting the typical
structure of the numberator and denominator graphs in the LF-MMI
function. For the numerator graph, we select the alignment graph from
the training set of WSJ with the largest number of states. This
alignment graph has 454 states and 1036 arcs. Then, we replicate this
graph 128 times to simulate a training mini-batch and we measure the
duration of the forward-backward on 128 input sequences of
pseudo-loglikelihoods. We set each sequence to have 700~frames as it
corresponds to the length of the largest likelihood sequence on WSJ.
For the denominator graph, we repeat the same procedure replacing the
alignment graph with a 3-gram phonotactic language model. The
denominator graph has 3022 states and 50984 arcs.

The duration of the different implementations are shown in
Table ~\ref{results:tab:bm}. First looking at the CPU version, we see
that the proposed Julia implementation significantly outperfoms the
logarithmic-domain PyChain implementation. The ``leaky-HMM''
version is drastically faster at the cost of yielding an approximate
result.  On GPU, the benefit of the ``leaky-HMM'' vanishes and the
implementation is even slower in the case of the numerator graph
benchmark. On the other  hand, our Julia implementation fully benefits
from the parallelism provided by the GPU device and shows more than
5 times speed up compared to logarithmic-domain PyChain implementation
on the denominator graph.
%
\begin{table}[ht]
\centering
\begin{tabular}{ccc}
    \toprule
    & \multicolumn{2}{c}{\textbf{Avg. duration (s)}} \\
    \textbf{System} & \textbf{LF-MMI} & \textbf{Neural network propagation} \\
    \toprule
    PyChain & 4.08 & 0.076  \\
    proposed & 0.220 & 0.180 \\
    \bottomrule
\end{tabular}
    \caption{Comparison of the average time spend in (i) the computation
    of the LF-MMI loss and its gradient, (ii) the forward and backward
    propagation through the neural network.}
\label{results:tab:time_analysis}
\end{table}

Next, we measure the overhead induced by the forward-backward algorithm
during the training. In Table \ref{results:tab:time_analysis}, we plot
the average duration of (i) the computation the LF-MMI loss and its
gradient as described in \eqref{eq:lfmmi_grad}, (ii) the forward and
backward propagation through the neural network computation excluding
the loss computation. These averages are estimated from 100
gradient computations during the WSJ training. We see that for the
Julia-based version the loss overhead is severely reduced and
accounts for only half the time of the total backpropagation.
Interestingly, one can see that KNet, the neural network backend used
in the Julia recipe, is actually slower than PyTorch, the neural
network backend used in the PyChain recipe. Therefore, the speed up of
the Julia-based training comes from our forward-backward algorithm
rather than a faster neural network toolkit.

\section{Conclusion and Future work}
\label{sec:future}
We have proposed a new implementation of the forward-backward algorithm
relying on semiring algebra. This implementation naturally fits in the
Julia language and achieves a significant speed up compared to a
competitive C++ implementation.

So far, we have only explored the use of the log-semifield,
however, it is trivial to extend our algorithm to other semirings.
Particulary, replacing the log-semiring with the tropical-semiring
would lead to a straightforward implementation of the Viterbi algorithm.
It is remarkable that with the advance programming languages, the
implementation of something as complex as a full-fledged speech decoder
can now be done in a few dozens lines of code.

Finally, we hope that this work will sparkle the interest of the speech
community to the rich capabilities offered by the Julia language
ecosystem.

\vfill
\pagebreak

\bibliographystyle{IEEEbib}
\bibliography{refs}

\end{document}